\documentclass[%
 reprint,
showpacs,
 amsmath,amssymb,
 aps,prl,
]{revtex4-1}

\usepackage{graphicx}
\usepackage{dcolumn}
\usepackage{bm}
\usepackage{color}

\begin{document}

\preprint{APS/123-QED}
\title{Thermodynamics of fluctuations based on time-and-space averages}

\author{James E. McClure} 
\affiliation{Virginia Polytechnic Institute \& State University, Blacksburg} 
\author{Steffen Berg}
\affiliation{Shell Global Solutions International B.V.
Grasweg 31,
1031HW Amsterdam,
The Netherlands}
\author{Ryan T. Armstrong}
\affiliation{University of New South Wales, Sydney}

\date{\today}

\begin{abstract}
We develop non-equilibrium theory by using averages in time and space as a generalized way to
upscale thermodynamics in non-ergodic systems. The approach offers a classical perspective on the energy dynamics in fluctuating systems. The rate of entropy production is shown to be explicitly scale dependent when considered in this context. We show that while any stationary process can be represented as having zero entropy production, second law constraints due to the Clausius theorem are preserved due to the fact that heat and
work are related based on conservation of energy. As a demonstration we consider the energy dynamics
for the Carnot cycle and for Maxwell's demon. We then consider non-stationary processes, applying time-and-space averages to characterize non-ergodic effects in heterogeneous systems where energy barriers such as compositional gradients are present. We show that the derived theory can be used to understand the origins of anomalous diffusion phenomena in systems where Fick's law applies at small length scales but not at large length scales. 
We then characterize fluctuations in capillary-dominated systems, which are non-stationary due to the irreversibility of cooperative events. 


\end{abstract}


            
\maketitle

\subsection*{Introduction}
The ergodic hypothesis is central to many results of statistical physics. The basic premise is that a system will explore all possible energetic micro-states if considered over a sufficiently long interval of time. The concept of ergodicity is thereby linked with the mixing of information within a system \cite{parry2004topics}.  
Canonical proofs of the ergodic hypothesis rely on the equivalence of spatial, temporal and ensemble averages in the limit of infinite time \cite{Neumann70,Birkhoff656}. Scale considerations and the rate of mixing necessarily constrain the applicability of the ergodic
hypothesis when considering finite regions of time, particularly for systems where
mixing is slow compared to the physical timescale of interest \cite{Palmer_1982}. 
Many physical systems are known to exhibit behavior that is inconsistent with the ergodic hypothesis.  
Well-known examples include anomalous diffusion in biological systems \cite{Weigel6438,Schwarzl_2017,Gnesotto_2018},
glassy systems \cite{Debenedetti_2001,Crisanti_2003,Tarjus_2005}, capillary phenomena and nucleation \cite{PhysRevLett.111.064501,PhysRevE.88.052116,CRANDALL2009574,PhysRevLett.87.055701,Iwamatsu_nucleation_2011,Winkler_etal_2019} and granular systems where multiscale effects are present due to jamming and force chains \cite{RevModPhys.90.015006,Herrmann1998,Mehta8244,Atia_etal_2018}. A common element for these systems is that they involve spatially heterogeneous materials where available thermal energy is insufficient to overcome internal energy barriers. This inhibits mixing and prevents the system from exploring all possible micro-states within the timescale of interest. 

\begin{figure}[ht]
\centering

\includegraphics[width=1.0\linewidth]{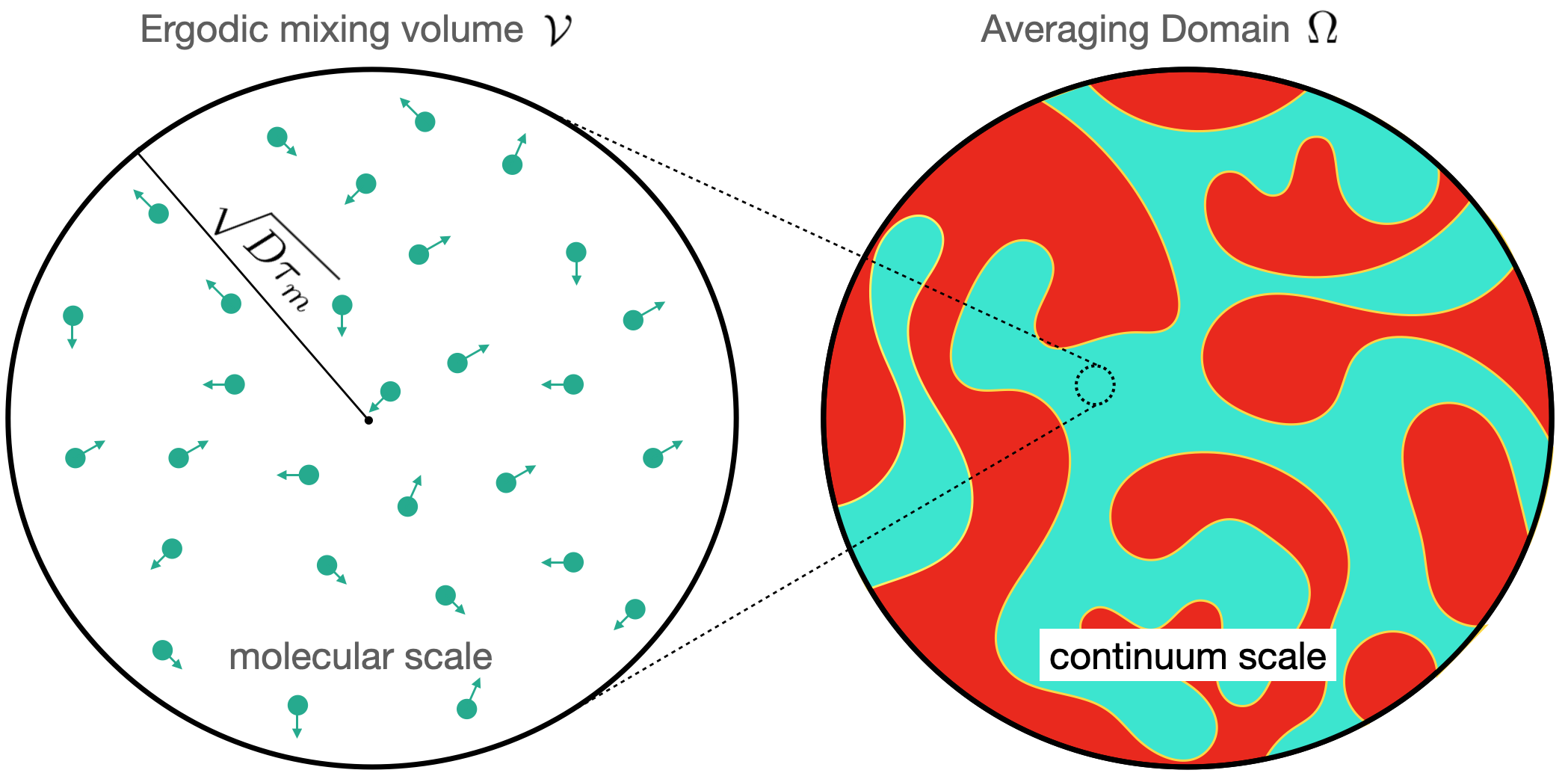}
\caption{The length scale for ergodicity can be estimated
based on $\sqrt{D\tau_m}$, the mean distance for diffusion over 
timescale $\tau_m$. We consider heterogeneous systems where the ergodic hypothesis holds at the scale of $\mathcal{V}$, but not at the larger scale of $\Omega$.}
\label{fig:spatial-scales}
\end{figure}

In this paper, we demonstrate that time-and-space averaging can be applied as a mechanism to mathematically mix information at the desired scale, providing a natural path forward in systems where non-ergodic effects are encountered.
Multiscale fluctuation terms arise in the non-equilibrium description due to spatial and temporal deviations associated with intensive thermodynamic variables. Our approach is rooted in classical thermodynamics and offers a formally distinct perspective on fluctuations as compared to statistical theory, e.g. \cite{Callen_1951,Kubo_1957,PhysRevLett.103.090601,Maes_2020,Wolpert_2020}. The presented methods 
can be a general tool for understanding how energy dynamics translate across length and time scales.
First we consider a basic example, demonstrating that the Carnot cycle and Maxwell's demon 
can be interpreted in terms of energy fluctuations. Then we consider mass transport phenomena in the context 
where Fick's law applies at a small length scale but fails at larger scales. Finally, the approach will be applied to non-equilibrium behavior in multiphase systems, where fluctuations occur due to capillary effects and confinement. How to characterize and interpret these fluctuations has been a long-standing problem for immiscible fluid flow in porous media, and has broad applications to other systems \cite{Morrow_1970,Pahlavan13780,Primkulov_2019,Cueto-Felgueroso_Juanes_2015,Armstrong_Ott_etal_2014,Berg_Ott_etal_13,Aryana_2013,Hassanizadeh_Gray_93,Bear_Nitao_1995,marle1981multiphase}.


\subsection*{Time-and-space averaged thermodynamics}

A classical thermodynamic description is defined by considering the internal energy to depend on 
the entropy $S$ as well as other extensive physical properties of the system, $X_i$ 
(e.g. volume, number of particles, etc.)
\cite{callen1960thermodynamics,Grmela_1997a,Grmela_1997b}
\begin{equation}
    U = U(S,X_1,X_2, \ldots, X_n) \;.
    \label{eq:internal-energy}
\end{equation}
Intensive quantities are then defined according to Euler's homogeneous function theorem,
\begin{equation}
    T = \Big( \frac{\partial{U}}{\partial S}\Big)_{X_i} \;, \quad  Y_i = \Big( \frac{\partial{U}}{\partial X_i}\Big)_{S,X_{j\neq i}}\;,
    \label{eq:euler-homogeneous}
\end{equation}
such that
\begin{equation}
    U = T S +   Y_i X_i
    \label{eq:euler}
\end{equation}
describes the internal energy of the system at equilibrium. 
In practice, measurements of thermodynamic quantities are 
averages carried out over some finite region of space and time. 
The measurement time $\tau_m$ can be used to infer
the size of the region where local equilibrium conditions exist.
For example, if a thermometer is used to measure the temperature,
any material that is close enough to the measurement point 
can be considered to be in local equilibrium with the measured value. 
Diffusive mechanisms are responsible for mixing energy within the system, since it
is the movement of the molecules and their interaction with each other that creates ergodic conditions.
Given a fixed timescale $\tau_m$, an associated length scale is obtained from the Einstein relation,
which predicts the mean squared displacement (MSD) for molecular trajectories $x(t)$,
\begin{equation}
\int_{0}^{\tau_m} x^2(t) d t \sim D \tau_m \;,
\label{eq:msd}
\end{equation}
where $D$ is the diffusion coefficient.
Since MSD predicts the average distance that molecules drift within time $\tau_m$,
the system is locally well-mixed at that length scale. 
Ergodic behavior should therefore be observed within a surrounding spherical region with volume
\begin{equation}
    \mathcal{V} \le  \frac{4\pi}{3} (D \tau_m)^{3/2}\;.
\end{equation}
During the elapsed time $\tau_m$, molecules will
explore a spatial region with size $\mathcal{V}$ such 
that spatial, temporal and ensemble averages are interchangeable at this scale. 
Smaller $\mathcal{V}$ can be chosen as long as the defined region is larger
than the molecular length scale. At the scale of $\mathcal{V}$, the theoretical bridge between the molecular 
and hydrodynamic description can be provided by statistical theory, relying on the validity of the
ergodic hypothesis \cite{Froemberg_2013}. 
Eq. \ref{eq:euler} can then be rescaled to treat extensive measures on a per-unit-volume basis, 
\begin{equation}
    \frac{U}{\mathcal{V}} = T \frac{S}{\mathcal{V}} +  Y_i \frac{X_i}{\mathcal{V}}\;.
    \label{eq:euler-per-unit-volume}
\end{equation}

\begin{figure*}[ht]
\centering
\includegraphics[width=0.9\linewidth]{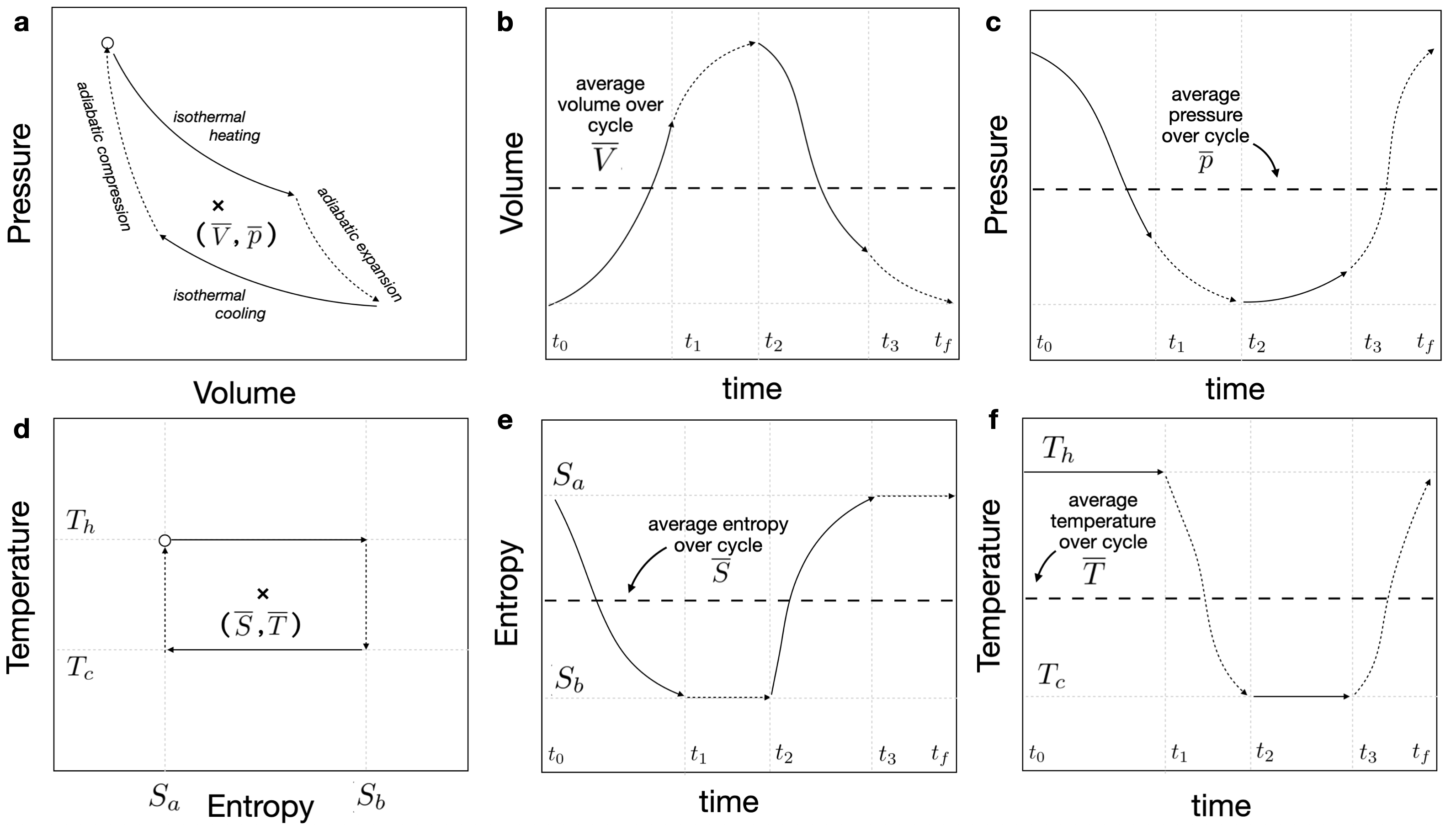}
\caption{Carnot cycle as a fluctuation: time averaging around the entire cycle 
result in constant average values of the entropy, $\overline{S}$ and temperature $\overline{T}$;
 volume $\overline{V}$ and pressure $\overline{p}$. For any cyclic process a stationary time series will be obtained for all thermodynamic quantities. Using a time average the energy dynamics can be embedded 
 within fluctuation terms that capture the net energy contribution over the cycle.}
\label{fig:carnot-fluctuation}
\end{figure*}

As depicted in Figure \ref{fig:spatial-scales}, we wish to develop theory that holds over some 
arbitrarily larger spatial region $\Omega$ (with volume $V>\mathcal{V}$),
and time interval $\Lambda$ (with duration $\lambda \ge \tau_m$). 
While the ergodic hypothesis must hold at the scale of $\mathcal{V}$, 
the system may be non-ergodic when considered at the larger scale of $\Omega$. 
Due to the potential failure of the ergodic hypothesis 
statistical mechanics is less well-suited to derive relevant hydrodynamic theory at  at larger scales. 
In this situation is desirable to define averages such that the ergodic hypothesis
applies at a small scale but not at larger scales. To define larger scale measures we apply the 
time-and-space averaging operator
\begin{equation}
    \big< f \big> \equiv\frac{1}{\lambda \mathcal{V}} \int_{\Lambda} \int_{\Omega} f dV dt\;.
    \label{eq:average}
\end{equation}
We note that this convention does not define an ensemble average, but instead
represents an explicit average over a region in time and space. The integral is 
therefore constructed to include the actual dynamics of the system, such
that insights into the system behavior can be inferred based on conservation of
energy with minimal assumptions.
Averages are defined such that extensive quantities retain the same physical meaning
across scales, 
\begin{equation}
    \overline{U} \equiv \big< U\big>\;, \quad \overline{S} \equiv \big< S \big>\;, \quad \overline{X}_i \equiv \big< X_i \big> \;.
    \label{eq:extensive}
\end{equation}
The intensive quantities are then defined as a weighted average with the conjugated 
extensive quantity from Eq. \ref{eq:extensive},
\begin{equation}
    \overline{T} \equiv \frac{\big< T S \big>}{\big< S \big>}\;, \quad 
    \overline{Y}_i \equiv \frac{\big< X_i Y_i\big>}{\big< X_i \big>} \;.
    \label{eq:intensive}
\end{equation}
These definitions ensure that the representation of the system energy is scale-consistent,
e.g., the product of entropy and temperature corresponds to the thermal energy \cite{Gray_Miller_14}.
Since entropy is additive, the temperature should be defined as the average thermal energy 
per unit of entropy. The averaged form is thus consistent with Eq. \ref{eq:euler},
\begin{equation}
    \overline{U} = \overline{T} \overline{S} +  \overline{X}_i \overline{Y}_i \;.
    \label{eq:average-euler}
\end{equation}

Non-equilibrium behavior can be considered by averaging the differential form of Eq. \ref{eq:euler}
\begin{eqnarray}
\frac{\partial\overline{U}}{\partial t} &=& \Big< T \frac{\partial S}{\partial t}\Big>
+  \Big< Y_i \frac{\partial X_i}{\partial t}\Big> 
\nonumber \\
   &=& \overline{T} \frac{\partial\overline{S}}{\partial t}  +
   \Big< (T - \overline{T})\frac{\partial S}{\partial t}\Big>
\nonumber \\ 
&& + \overline{Y}_i \frac{\partial\overline{X}_i}{\partial t} 
+ \Big< (Y_i - \overline{Y}_i) \frac{\partial X_i}{\partial t}\Big> \;.
\label{eq:non-eq-i}
\end{eqnarray}
We now define multiscale deviation terms as 
\begin{equation}
   {T}^\prime \equiv  T-\overline{T}\;, \quad  {Y}_i^\prime \equiv Y_i-\overline{Y}_i \;.
\end{equation}
Using the definitions from Eqs. \ref{eq:extensive} and \ref{eq:intensive}
we can show that
\begin{equation}
    \Big< T^\prime \frac{\partial S}{\partial t}\Big> = 
    -\Big< S \frac{\partial T^\prime}{\partial t}\Big> \;, \quad
        \Big< Y_i^\prime \frac{\partial X_i}{\partial t}\Big> = 
    -\Big< X_i \frac{\partial Y_i^\prime}{\partial t}\Big> \;.
\end{equation}
which can be substituted into Eq. \ref{eq:non-eq-i} and rearranged to obtain
an entropy inequality
\begin{eqnarray}
   \frac{\partial\overline{S}}{\partial t}  &=& \frac {1}{\overline{T}} \Big[
   \frac{\partial\overline{U}}{\partial t}
    - \overline{Y}_i \frac{\partial\overline{X}_i}{\partial t}  
  +\underbrace{\Big< S \frac{\partial{T}^\prime}{\partial t}\Big>
    + \Big< X_i \frac{\partial{Y}_i^\prime}{\partial t}\Big> }_{\mbox{fluctuation terms}} 
    \Big] \ge 0 \;.
        \nonumber \\ && 
    \label{eq:entropy-inequality}
\end{eqnarray}
The result is easily recognizable as the fundamental relation of non-equilibrium thermodynamics
(e.g. \cite{degroot_Mazur_2013}), but with additional terms associated with fluctuations.
The fluctuations contribute to the energy dynamics whenever intensive variables deviate non-linearly
from their average values within the spatial region $\Omega$ and time interval $\Lambda$. 

\subsection*{The Carnot Cycle as a Fluctuation}

Spatial averages have been explored extensively in the context of classical non-equilibrium thermodynamics 
\cite{Gray_Miller_14}. Extending this approach to also include time averages provides a way to 
smooth the temporal dynamics of the system. To illustrate how this works, we consider the familiar
example of the Carnot cycle, as depicted in Figure \ref{fig:carnot-fluctuation}. 
The cycle begins at time $t_0$ and completes at time $t_f$, repeating in periodic fashion. 
The time average does not depend on the region of time $\Lambda$ as long as the duration 
$\lambda$ is an integer multiple of the period $t_f - t_0$. We choose 
$\lambda = t_f - t_0$ so that $1 / \lambda$ corresponds to the cycle frequency. 
For the Carnot cycle the thermodynamic state is described 
by $U(S,V)$. The non-equilibrium behavior is described in an average form 
according to Eq. \ref{eq:entropy-inequality},
\begin{eqnarray}
   \frac{\partial\overline{U}}{\partial t}
   -\overline{T} \frac{\partial\overline{S}}{\partial t}  
    + \overline{p} \frac{\partial\overline{V}}{\partial t}  
  +\Big< S \frac{\partial{T}^\prime}{\partial t}\Big>
    - \Big< V \frac{\partial{p}^\prime}{\partial t}\Big>  = 0 \;,
    \label{eq:time-avg-non-eq}
\end{eqnarray}
where $\overline{U}$, $\overline{S}$, $\overline{V}$, $\overline{T}$ and $\overline{p}$
are time-and-space average values for the complete cycle. Since the associated time series is stationary,
these are each constant when the time averaging interval is an integer multiple of $\lambda = t_f - t_0$. 
This means that 
\begin{equation}
    \frac{\partial\overline{U}}{\partial t} = 0\;, \quad 
    \frac{\partial\overline{S}}{\partial t} = 0\;, \quad 
    \frac{\partial\overline{V}}{\partial t} = 0\;.
    \label{eq:extensive-cycle}
\end{equation}
In other words, $\overline{S}$, $\overline{V}$, $\overline{T}$ and $\overline{p}$
are the average values around which the system is fluctuating, as shown in Fig. \ref{fig:carnot-fluctuation}a
--f. The energy dynamics are fully described by the relationship
between the fluctuations,
\begin{eqnarray}
\Big< V \frac{\partial{p}^\prime}{\partial t}\Big> =  \Big< S \frac{\partial{T}^\prime}{\partial t}\Big>\;.
    \label{eq:time-avg-fluctuation}
\end{eqnarray}
We now separately consider each fluctuation. First,
the pressure fluctuation is identical to the rate of pressure volume work
$W$ (with the sign convention chosen to obtain positive external work), 
\begin{eqnarray}
    \Big< V \frac{\partial{p}^\prime}{\partial t}\Big> 
    &=&  -\Big< p \frac{\partial{V}}{\partial t}\Big> 
     =  \Big< \frac{\partial W}{\partial t} \Big> \;.
\end{eqnarray}
This is a consequence of the fact that the averages $\overline{p}$ and $\overline{V}$ are each constant. The pressure fluctuation therefore directly corresponds to the power output. 
Next we treat the temperature fluctuation.  For the Carnot cycle all temperature changes occur 
during isentropic conditions. Using the additive property for the time integral, it is 
straightforward to show that
\begin{eqnarray}
\Big< S \frac{\partial{T}^\prime}{\partial t}\Big> 
 &=& \Big< S \frac{\partial{T}}{\partial t}\Big>  \nonumber \\
 &=&  \frac{1}{\lambda} \Bigg(\int_{t_1}^{t_2} S_a \frac{\partial T}{\partial t} dt + 
        \int_{t_3}^{t_f} S_b \frac{\partial T}{\partial t} dt \Bigg)\nonumber \\ 
       &=& \frac{1}{\lambda} (S_b - S_a) (T_h - T_c) \;.
\end{eqnarray}
Inserting these into Eq. \ref{eq:time-avg-non-eq} and using the fact that the heat added to the
system is given by $Q_h = T_h (S_b-S_a)$ we obtain
\begin{eqnarray}
\Big< \frac{\partial W}{\partial t} \Big>
     =  \frac{1}{\lambda} \frac{Q_h (T_h - T_c)}{T_h} \;.
    \label{eq:time-avg-non-eq-carnot-i}
\end{eqnarray}
This is identical to the standard result for the efficiency of a Carnot engine, but is
formulated as an expression for the power output based on the cycle frequency $\lambda$. 
This demonstrates that in a time averaged formulation, any energy dynamics that occur faster 
than the duration for the time averaging window will be re-cast as fluctuations. Since the integral
defined by Eq. \ref{eq:average} explicitly accounts for the path of the system, the energy dynamics are 
fully captured by the homogenized representation. With respect to the second law of thermodynamics,
a critical insight from this exercise is that the interpretation of the entropy is scale dependent,
since the rate of entropy production depends explicitly on the timescale at which a process is considered. 
This will be particularly important in heterogeneous systems where the are crossover times due to the
scaling behavior for dominant physical processes.
For any stationary process the rate of entropy production will be zero if the dynamics are considered over 
a sufficiently long timescale. We note that this does not contradict the Clausius theorem, since the relationship between 
the rate of of heat added and the rate of work is unchanged based on the time average. In other words a scale-consistent representation of the energy dynamics is recovered. 

\subsection*{Fluctuations of Maxwell's demon}

\begin{figure}[ht]
\centering
\includegraphics[width=1.0\linewidth]{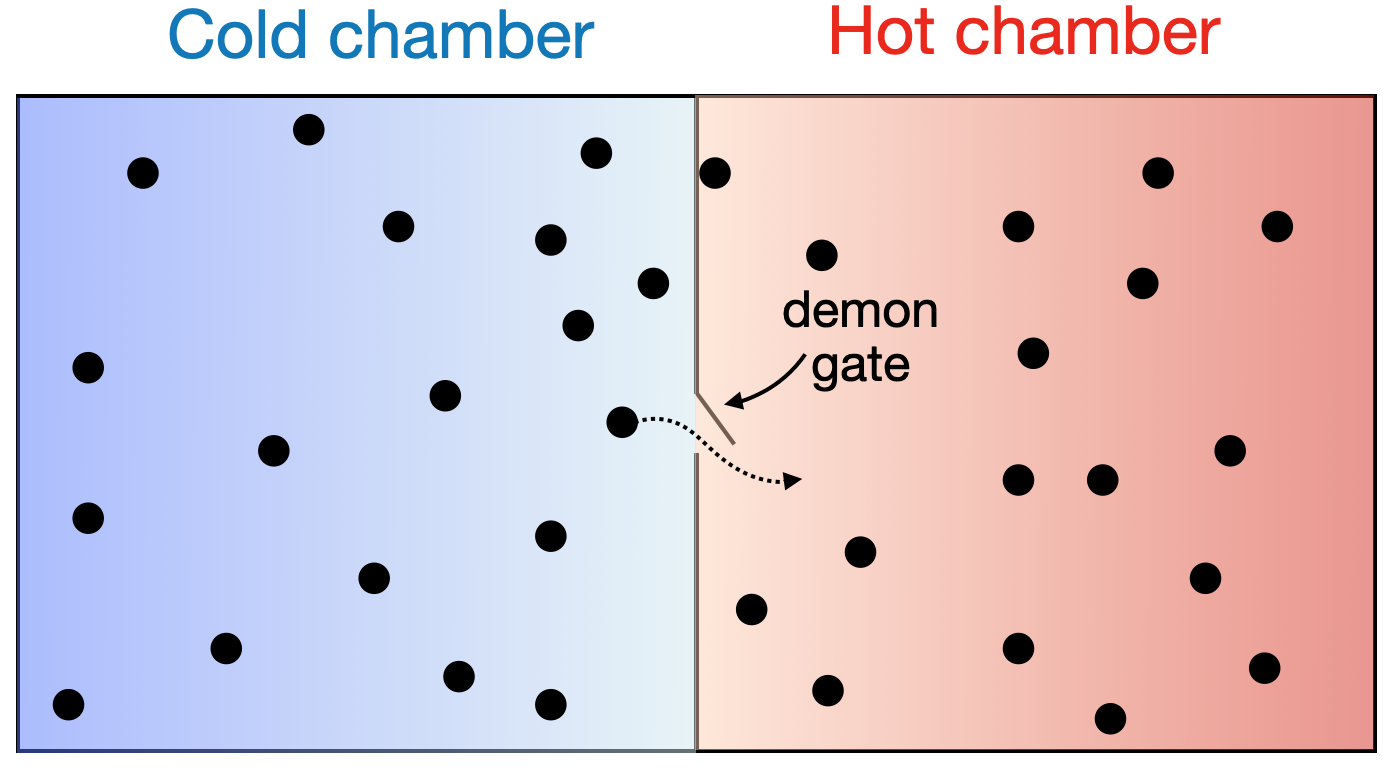}
\caption{``Maxwell's demon" considers a thought experiment in which a demon
selectively permits the migration of hot molecules from one chamber to the other. 
Due to the heat flux carried by the transmitted molecules, the temperature of the hot
chamber increases, leading to a thermal gradient between the 
chambers. The demon's actions are constrained by a symmetry law relating thermal fluctuations. }
\label{fig:maxwell-demon}
\end{figure}

To further illustrate the role of fluctuations, we consider
the actions of Maxwell's demon based on the two-chamber system shown in Figure \ref{fig:maxwell-demon} \cite{maxwell_1871,maxwell_1879}. The demon operates a gate, selectively allowing fast-moving molecules to move from the cold chamber to the hot chamber. In apparent violation of thermodynamic intuition, the associated transfer of kinetic energy increases the temperature of the hot chamber. Maxwell's demon can be considered in the context of non-ergodic behavior, as the demon controls the mixing of information between the two chambers. The demon defines 
locally non-stationary behavior by applying molecular scale rules that prevent
prevent a forward process from reaching equilibrium with the corresponding reverse process. 
However, this does not mean the process is irreversible; it only means that the equilibrium has been
delayed. We will show that if a conservative demon is disabled, the system will return to it's original state.
Fluctuations are an appropriate tool for understanding the energy dynamics in such a system.
Furthermore, the system is spatially heterogeneous, and the chambers can be denoted as sub-regions of the system, 
$\Omega_{h}$ and $\Omega_{c}$. Based on these definitions a discrete aspect is introduced into the 
system representation, since the crossing of molecules from one sub-region to the other occur as 
discrete events. Time averaging smooths the effect of these crossings such that the action of the demon can be modeled as being continuous with respect to time. 

For an ideal monatomic gas the entropy is given by the Sackur-Tetrode equation \cite{Sackur_1911,Tetrode_1912,Ben-Naim_2008}
\begin{equation}
S(U,V,N) = k_B N \Bigg[ \frac 52 + \ln  \frac VN  
    +\frac 32 \ln \frac U N + \frac 32 \ln \frac{2\pi m}{h^2}
\Bigg] \;,
\label{eq:sackur-tetrode}
\end{equation}
where $m$ is the particle mass and $h$ is Planck's constant.
The expression for entropy as a state function is sufficient to determine the form
$U(S,V,N)$. The intensive quantities can be determined directly from their 
thermodynamic definitions
\begin{eqnarray}
  T \equiv \Bigg( \frac{\partial U}{\partial S} \Bigg)_{N,V} 
  &=& \frac{2U}{3 k_B N}\;,
    \nonumber \\
  p \equiv -\Bigg( \frac{\partial U}{\partial V} \Bigg)_{N,S} 
  &=& \frac{k_B N T}{V} \;,
   \nonumber  \\
  \mu \equiv \Bigg( \frac{\partial U}{\partial N} \Bigg)_{S,V} 
    &=& -k_B T \Big(\ln \frac VN + \frac 32 \frac U N + \frac 32 \ln \frac{2\pi m}{h^2}\Big) \;.
   \nonumber \\
   \label{eq:ideal-intensive}
\end{eqnarray}
Noting that $U=\frac 32 k_B N T$ for an ideal gas, it is easy to show these forms
are consistent with Eq. \ref{eq:euler-homogeneous} in the particular form $U=TS-pV+\mu N$.
To treat the demon we must subdivide the system based on thermodynamics within each 
chamber. This is accomplished by sub-setting the system into regions
based on the indicator function $\Upsilon_i$
\begin{equation}
    \Upsilon_i (\mathbf{x}) = \left \{  \begin{array}{cc} 1& \mbox{if $\mathbf{x} \in \Omega_i$} \\
    0 & \mbox{otherwise} \end{array}\right.
    \label{eq:subset}
\end{equation}
for $i \in \{h,c\}$. Consistent with Eqs. \ref{eq:extensive} and \ref{eq:intensive}, separate
temperature, entropy, pressure, chemical potential and number of molecules
are obtained for each chamber.
For extensive properties, fields are constructed based on the subset operation,
\begin{equation}
    \frac{S_i}{\mathcal V} \equiv \frac{S \Upsilon_i}{\mathcal{V}} \;, \quad 
    \frac{N_i}{\mathcal V} \equiv \frac{N \Upsilon_i}{\mathcal{V}} \;, \quad 
    V_i = \mathcal{V} \Upsilon_i \;,
\end{equation}
where $\mathcal{V}$ has been included explicitly in the definitions to account for the 
fact that the reference volume used to define fields should not be infinitely small
(i.e. should  exceed the molecular length scale). Averages are then defined as
time-and-space averages
\begin{eqnarray}
\overline{S}_i \equiv \big< S_i\big>\;, \quad
\overline{T}_i \equiv  \frac{\big< T S_i\big>}{\big< S_i \big>} \;,
\nonumber \\
\overline{V}_{i} \equiv \big< V_i \big>\;, \quad
\overline{p}_{i} \equiv  \frac{\big< p V_i \big>}{\big< V_i  \big>} \;,
\nonumber \\
\overline{N}_{i} \equiv \big< N_i \big>\;, \quad
\overline{\mu}_{i} \equiv  \frac{\big< \mu N_i \big>}{\big< N_i  \big>} \;.
\end{eqnarray}
For simplicity we consider the case where volume for each chamber is constant, with
${V}_{c} = {V}_{h}$. These definitions ensure that the entropy is additive, 
\begin{equation}
    \overline{S} =  \overline{S}_c +  \overline{S}_h \;,
    \label{eq:maxwell-total-entropy}
 \end{equation}
and that the total thermal energy is conserved based on the definition of the temperature,
\begin{eqnarray}
    \overline{T} \overline{S} 
     &=&  \overline{T}_c \overline{S}_c +  \overline{T}_h \overline{S}_h \;.
\end{eqnarray}
For the number of molecules and chemical potential,
\begin{eqnarray}
    \overline{\mu}  \overline{N} 
     &=&  \overline{\mu}_{c} \overline{N}_{c} +  \overline{\mu}_{h} \overline{N}_{h} \;.
\end{eqnarray}
It is clear that generic thermodynamic quantities can be defined by sub-setting
a heterogeneous system into regions. The result of the sub-setting operation is 
functionally equivalent to an alternative specification of the extensive variables,
\begin{equation}
    U = U(S_h, S_c, V_h, V_c, N_h, N_c) \;.
    \label{eq:maxwell-subdivision}
\end{equation}
However, since the set operation defined from $\Upsilon_i$ introduces a discrete element into the system representation, 
apparent discontinuities can result when considering energy exchanges between the entities. 
Averaging in time removes these discontinuities to ensure a smooth representation for the dynamics
within sub-regions. As an example, even though the Hamiltonian for the molecular system is continuous, the molecular crossing shown in Figure \ref{fig:maxwell-demon} is a discrete event. Combining the sub-set operation with
a time average leads to a corresponding fluctuation theorem because any energy gained by the hot chamber is directly
lost by the cold chamber. The fluctuation terms impose a symmetry constraint on the demon's action on the basis 
that the demon must conserve mass and energy. 

Since the system is closed the total number of particles is conserved,
\begin{equation}
    \frac{\partial\overline{N}_{h}}{\partial t}  = - \frac{\partial\overline{N}_{c}}{\partial t} \;.
\end{equation}
Subdivision of the system into hot and cold compartments leads to
the following fluctuation constraint based on conservation of energy,
\begin{eqnarray}
\frac{\partial\overline{S}}{\partial t} 
  &=& \frac{1}{\overline{T}}\sum_{i} \Bigg[
 \Big< S_i \frac{\partial{T_i}^\prime}{\partial t}\Big>
 - \Big< V_i \frac{\partial{p_i}^\prime}{\partial t}\Big>
 + \Big< N_{i} \frac{\partial{\mu}_{i}^\prime}{\partial t}\Big> \Bigg] 
    \label{eq:entropy-inequality-demon-ii}
\nonumber \\
&& +\Big( \frac{\overline{\mu}_{h} - \overline{\mu}_{c}}{T}\Big)\frac{\partial\overline{N}_{c}}{\partial t}     \;,
\end{eqnarray}
where  $T_i^\prime = T_i - \overline{T}$, $\mu_i^\prime = \mu_i - \overline{\mu}$, and
$p_i^\prime = p_i - \overline{p}$ on $\Omega_i$ with
$i\in\{c,h\}$. 

\begin{figure*}[ht]
\centering
\includegraphics[width=0.95\linewidth]{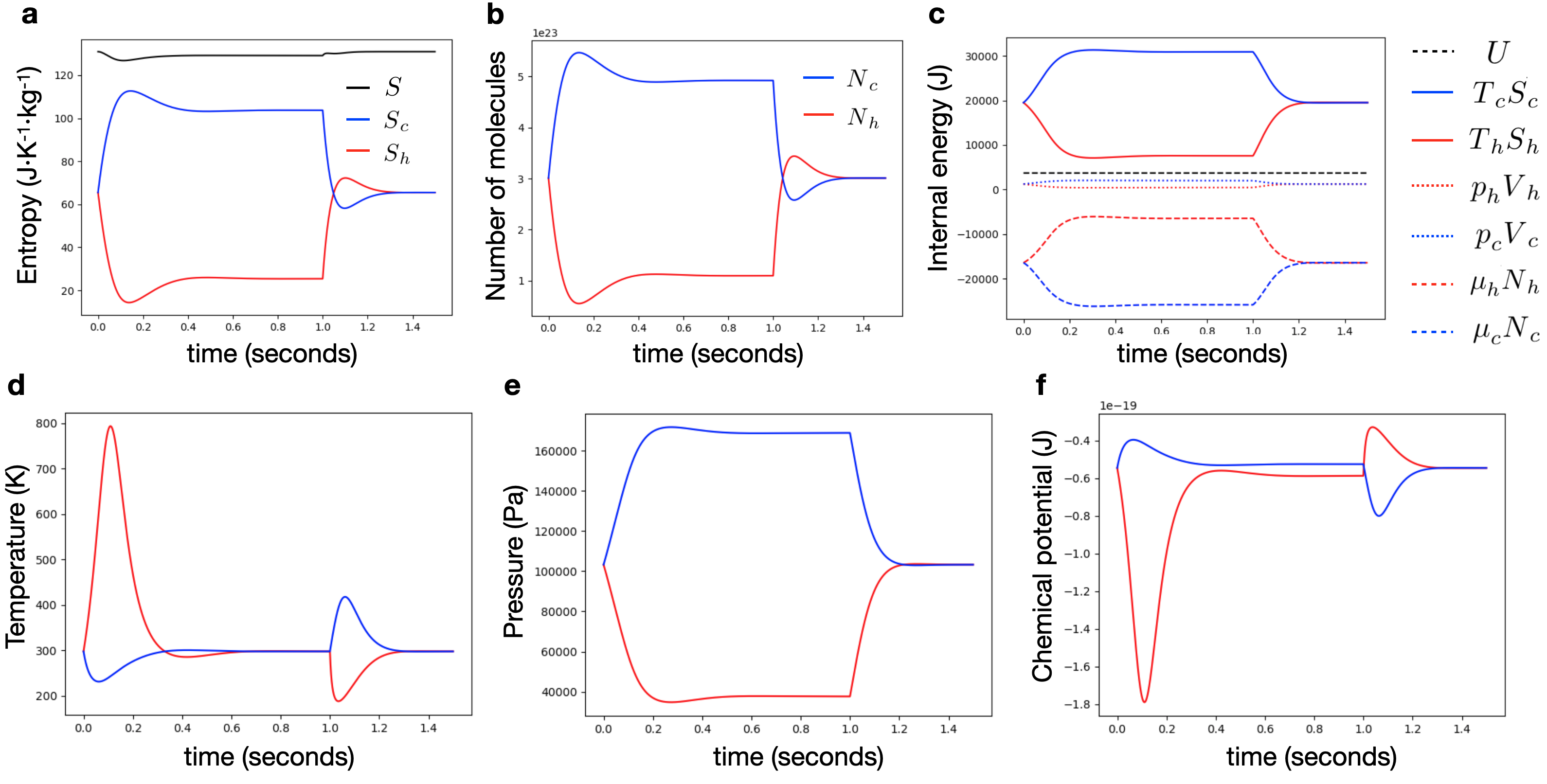}
\caption{Fluctuations due to operation of a conservative Maxwell's demon: the demon is activated
at $t=0$ second and de-activated at time $t=1.0$ second to obtain a stationary time series. 
Fluctuations to the intensive properties are observed based on the redistribution of mass and energy within the system.}
\label{fig:demon-fluctuation}
\end{figure*}

It has been argued from an information theory perspective that the demon must be able to perform 
measurements in order to function as described by Maxwell \cite{Earman_Norton_1999}. 
For example, the demon would need to know the temperature of the hot chamber so that it 
could determine which molecules should pass through the gate. However, a simple thought 
experiment demonstrates that a demon can generate gradients without relying on any non-local information
or altering the energy for any molecules that it comes into contact with.
Consider a demon that is tuned to operate at some particular temperature $T_{d}$. The demon will allow any molecule with speed greater than one standard deviation above the mean for the Maxwell distribution of speeds to pass from the cold chamber to the hot chamber, i.e. molecules with speed
\begin{equation}
    v_{d} \ge \sqrt{\frac{3 k_B T_d}{m}} \;.
    \label{eq:demon-rule}
\end{equation}
The statistics for the crossings can be determined from the Maxwell distribution.
If the demon operates at temperatures in the vicinity of $T_d$, it can drive the formation of a
gradient without any knowledge of the system state. The demon only needs to measure the molecular 
speed, which may be done locally and reversibly without retaining any memory of the measurement. 
Of course, this action cannot be performed without a constraint: molecules with sufficient 
speed must hit the portion of the wall where the demon gate is located, since the demon cannot 
impose any forces to attract particles toward its location. The probability for fast molecules to 
hit the demon gate depends on $T_c$ based on the Maxwell distribution of speeds and also on the area of 
the gate relative to the area of the partition between the two chambers. As $T_c$ decreases, so too will 
the rate at which sufficiently energetic molecules hit the gate. With the demon's behavior defined
statistically based on Eq. \ref{eq:demon-rule} we can calculate the rate for mass exchange.
The number of particles moving from the cold chamber to the hot chamber is 
\begin{eqnarray}
    \frac{\partial {N}_{ch}}{\partial t}  &=& \frac {N_c A_g}{V_c} \int_{v_d}^{\infty} \sqrt{\frac{m}{2\pi k_B T_c}} s e^{-m s^2 / 2 k_B T_c} ds \;, \nonumber  \\
   &=& - \frac{N_c A_g}{ V_c }\sqrt{\frac{k_B T_c}{2 \pi m}} e^{- m v_d^2 / 2 k_B T_c}  \;. 
    \label{eq:demon-rule-mass}
\end{eqnarray}
where $A_g$ is the area of the gate between
chambers. Similarly, the rate of energy exchange is 
\begin{eqnarray}
    \frac{\partial{U}_{ch}}{\partial t} &=& \frac {N_c A_g}{V_c} \int_{v_d}^{\infty} 
    \sqrt{\frac{m}{2\pi k_B T_c}} \frac{ms^3}{2} e^{-m s^2 / 2 k_B T_c} ds \;,
     \\
     &=&  - \frac {m N_c A_g}{ V_c \sqrt{2\pi}} \Bigg( \frac{k_B T_c}{m} \Bigg)^{3/2}
    \Bigg(1 + \frac{m v_d^2}{2 k_B T_c}\Bigg)  e^{-mv_d^2 /2 k_B T_c }  \;. \nonumber
    \label{eq:demon-rule-energy}
\end{eqnarray}
We consider a two-way valve so that a steady-state can be achieved based on
the redistribution of slower molecules to the cold chamber. 
A simple way to achieve this is to allow any molecule that hits 
the gate to travel back to the cold chamber. This clearly requires no measurement
and meets the constraints used previously. The statistics for the
particles gained by the cold chamber are given by
\begin{eqnarray}
    \frac{\partial {N}_{hc}}{\partial t}  
   &=& \frac{N_h A_g}{ V_h }\sqrt{\frac{k_B T_h}{2 \pi m}}\;,    \\
    \frac{\partial {U}_{hc}}{\partial t}  
     &=&  \frac {m N_h A_g}{\sqrt{2\pi} V_h} \Bigg( \frac{k_B T_h}{m} \Bigg)^{3/2} \;. 
    \label{eq:demon-rule-energy}
\end{eqnarray}
Since the rules used to define this
behavior do not require non-local information, the demon will be able to 
spatially segregate particles based on conservation of energy and without violating physical laws. Fast particles will accumulate in the hot
chamber, and slow particles will accumulate in the cold chamber. The primary mechanism needed to
achieve this is to prevent slow molecules from entering the hot chamber. The role played by 
the demon is therefore to delay equilibrium by preventing energy from partitioning itself equally
between the two chambers. Compared to the initial state, the forward process (particles move from
hot to cold) cannot reach an equilibrium with the reverse process (particles move from cold to hot).
This produces a non-stationary process. However, since no energy is removed or added, stationary behavior can be restored by simply disabling the demon. 


We now treat the fluctuations that are created due to operation of the hypothetical demon. Based on Eqs. \ref{eq:demon-rule-mass} and \ref{eq:demon-rule-energy}  
the total mass and energy exchange between chambers is defined for each
timestep $\delta t$
\begin{eqnarray}
  \delta N_c &=&\Bigg[ \frac{N_h A_g}{ V_h }\sqrt{\frac{k_B T_h}{2 \pi m}} - \frac{N_c A_g}{ V_c }\sqrt{\frac{k_B T_c}{2 \pi m}} e^{- m v_d^2 / 2 k_B T_c} 
    \Bigg]\delta t \;, \nonumber \\ 
  \delta U_c &=& 
  \frac {m A_g}{\sqrt{2\pi} V_h} \Bigg[ N_h \Bigg( \frac{k_B T_h}{m} \Bigg)^{3/2} 
    \nonumber \\ 
 &&    - N_c \Bigg( \frac{k_B T_c}{m} \Bigg)^{3/2} 
  \Bigg(1 + \frac{m v_d^2}{2 k_B T_c}\Bigg)  e^{-mv_d^2 /2 k_B T_c } \Bigg]\delta t \;, 
  \nonumber \\
\end{eqnarray}
The energy and particle number are updated so that conservation is strictly obeyed
\begin{eqnarray}
N_c(t+\delta t) &\leftarrow& N_c(t) + \delta N_c \;,
\nonumber \\
N_h(t+\delta t) &\leftarrow& N_h(t)  - \delta N_c \;,
\nonumber \\
U_c(t+\delta t) &\leftarrow& U_c(t)  + \delta U_c \;, 
\nonumber \\
U_h(t+\delta t) &\leftarrow& U_h(t)  - \delta U_c \;.
\label{eq:demon-update-rule}
\end{eqnarray}
The system is closed by assuming that Eqs. \ref{eq:sackur-tetrode} and \ref{eq:ideal-intensive}
hold separately within each sub-region. 
The initial condition sets equilibrium conditions in each chamber with 1 mol of Helium 
atoms equally divided between the two chambers at $T_d=298$K. The demon is activated at $t=0$
with $\delta t = 0.00025$ sec. The demon operates until $t = 1.0$ second, ultimately 
reaching a steady-state. The difference in pressures is accounted for by the differential
momentum transfer that results from the demons operation according to Eqs. \ref{eq:demon-rule-mass} -- \ref{eq:demon-update-rule}. Once steady-state is achieved the demon is disabled, and the system returns exactly 
to its initial state and producing a stationary time series. 
For the results shown in Fig. \ref{fig:demon-fluctuation}, the system is explicitly ergodic by construction, since the Maxwell distribution has been assumed (meaning equipartition of energy is strictly
observed within each chamber). 
At this fast timescale entropy production occurs based on the redistribution of mass an energy between the two chambers (see Fig. \ref{fig:demon-fluctuation}a). 
We note that molecular rules do not necessarily need to satisfy positive entropy production, since entropy is a statistical concept that does not apply at the deterministic scale of an individual molecule. 
However, since no heat is added to the two-chamber system, the Clausius theorem remains valid. Unlike the Carnot engine, the demon does not require added heat or work to generate the fluctuation. The demon is simply delaying the equilibrium of the system. Since no energy has been removed from the system, it can be treated as a reversible process when considered
over longer time intervals.

If the timescale for averaging is small,
fluctuations to intensive properties will be zero based on the
fact that rate of change is locally linear over short timescales. 
Fluctuation terms contribute when considering behavior over long timescales.
At the end of the cycle, the entropy and particle number returns exactly to their original value, meaning that 
\begin{eqnarray}
    \frac{\partial \overline{S}}{\partial t} = 0 \;,  \quad \frac{\partial \overline{N}_c}{\partial t} = 0.
    \label{eq:demon-reversible}
\end{eqnarray}
In other words, zero entropy production is observed when the system is considered over the longer timescale. 
Inserting Eq. \ref{eq:demon-reversible} into Eq. \ref{eq:entropy-inequality-demon-ii} 
leads to fluctuation criterion at long times
\begin{eqnarray}
 \Big< S_i \frac{\partial{T_i}^\prime}{\partial t}\Big>
 - \Big< V_i \frac{\partial{p_i}^\prime}{\partial t}\Big>
 + \Big< N_{i} \frac{\partial{\mu}_{i}^\prime}{\partial t}\Big> = 0   \;.
 \label{eq:demon-fluctuation}
\end{eqnarray}
This result simply means that thermal fluctuations must obey 
conservation of energy, consistent with results established 
from the perspective of microscopic reversibility \cite{Noether_1918,Noether_1971}.
The result is a simple fluctuation theorem that is distinct from approaches proposed by other treatments \cite{Touchette_2000,Sagawa_2010,Shiraishi_2015,Koski_2014,Rupprecht_2019,Rupprecht_2020}.
A basic challenge presented by Eq. \ref{eq:demon-fluctuation} is that the fluctuation criterion
depends on the total entropy, since this is needed to compute $\Big<S_i \frac{\partial T_i^\prime}{\partial t} \Big>$. 
While the Sackur-Tetrode equation provides an adequate approximation for an ideal gas, the situation
for thermal fluctuations is not easily generalized. The demon illustrates two important features of 
systems where multi-scale fluctuations are important. First, it defines an energy barrier that prevents mixing between
the hot and cold chambers. Second, the problem is associated with gradients in an intensive 
thermodynamic property. We now demonstrate the importance of these 
terms in capillary-dominated systems, where gradients in composition and chemical potential lead to length scale heterogeneity.

\subsection*{Mass diffusion}

The advantages of time-and-space averaging are particularly intriguing for spatially
heterogeneous systems. In this section we consider the application to mass diffusion, 
which is a common source of spatial heterogeneity. 
Compositional gradients are a 
key feature in these systems, and these heterogeneities cause fluctuations when the 
system is stimulated. In such cases the fluctuation terms can be directly linked to classical microscopic non-equilibrium thermodynamics using phenomenological equations established based on the theory of Onsager \cite{Onsager_1931a}. Phenomenological equations provide the basis to establish the rates at which particular processes occur, linking the temporal, spatial and energy scales. Fick's law can be directly recovered at length scales that are accessible based on molecular dynamics simulations \cite{XU2006452}. In the context of Onsager, phenomenological equations are derived based on a local near-equilibrium assumption. The associated reciprocal relations can be derived based on an assumption of microscopic reversibility. These conditions are presumed to hold at the scale of $\mathcal{V}$, since this scale is defined based on the local diffusion coefficient. The relevant linear phenomenological equation is Fick's law, which is assumed to describe the non-equilibrium behavior at the small scale of $\mathcal{V}$,
\begin{equation}
    \frac{\partial {\rho_k}}{\partial t} - \nabla \cdot ( \sf{D}_k \cdot \nabla \mu_k) = 0\;,
        \label{eq:Fick}
\end{equation}
where $\sf{D}_k$ is the mass diffusion tensor, $\rho_k$ is the density and $\mu_k$ is chemical potential.
Fick's law asserts that there is a locally linear non-equilibrium response to compositional gradients.
However, we note that in a heterogeneous system linear response theory will fail at larger scales due to
non-linearity in the composition, diffusion coefficient and chemical potential.
Such non-linearities will inevitably lead to anomalous diffusion phenomena, and associated non-equilibrium behaviors can be understood in terms of multi-scale fluctuations.

Standard non-equilibrium treatment for mass diffusion is obtained by considering the internal energy to be described by 
$U(S,N_k)$, with $N_k$ being the number of molecules 
of component $k$. 
For a closed system Eq. \ref{eq:entropy-inequality} can be written as
\begin{equation}
       \frac{\partial\overline{S}}{\partial t} = \frac {1}{\overline{T}} 
       \Bigg[
 \Big<  S \frac{\partial{T}^\prime}{\partial t}\Big>
 - \overline{\mu}_k \frac{\partial\overline{N}_k}{\partial t}      
 + \Big< N_k \frac{\partial{\mu}_k^\prime}{\partial t}\Big> 
    \Bigg] \ge 0 \;.
    \label{eq:entropy-inequality-mass}
\end{equation}
In an isothermal system the temperature fluctuation is zero.  The remaining
terms can be directly interpreted based on Fick's law,
\begin{eqnarray}
\overline{\mu}_k \frac{\partial\overline{N}_k}{\partial t}      
 -  \Big< N_k \frac{\partial{\mu}_k^\prime}{\partial t}\Big>  
 &=&
      \overline{\mu}_k \frac{\partial\overline{N}_k}{\partial t}  
     -\Big < N_k \frac{\partial (\mu_k-\overline{\mu}_k)}{\partial t}\Big>
     \nonumber \\
    &=&  
    \overline{\mu}_k \frac{\partial\overline{N}_k}{\partial t}  
    + \overline{N}_k \frac{\partial \overline{\mu}_k }{\partial t}
     -\Big < N_k \frac{\partial \mu_k}{\partial t}\Big>
         \nonumber \\
    &=&  
   \frac{\partial (\overline{\mu}_k \overline{N}_k  ) }{\partial t} 
   -\frac{\partial  \big<\mu_k N_k  \big> }{\partial t}
     + \Big< \mu_k \frac{\partial N_k}{\partial t}\Big> \;.
         \nonumber 
 \end{eqnarray}
The first two terms cancel based on the fact that 
$\overline{\mu}_k \overline{N}_k =\big< \mu_k  N_k  \big>$ according to
the definition given in Eq. \ref{eq:intensive}. Now assuming that Fick's
law holds at the microscopic scale with $\rho_k = N_k/ \mathcal{V}$, 
we arrive at 
\begin{eqnarray}
\frac{1}{\mathcal{V}}  \Big< \mu_k \frac{\partial N_k}{\partial t}\Big>
 &=&
    \Big< \mu_k \nabla \cdot ( \sf{D}_k \cdot\nabla \mu_k) \Big >\;.
    \label{eq:Fick-i}
\end{eqnarray}
Inserting this into Eq. \ref{eq:entropy-inequality-mass} it is evident that the dissipation is entirely determined from the contribution of the spatial gradients. In other words, the length
scale associated with gradients is fundamentally linked with the timescale for energy dissipation based on the phenomenological coefficient. Furthermore, assuming Fick's law at the microscopic scale does not imply it will hold at larger scales. 
To see this we can formally average the right-hand
side of Eq. \ref{eq:Fick-i},
\begin{eqnarray}
    \Big< \mu_k \nabla \cdot ( \sf{D}_k \cdot\nabla \mu_k) \Big > 
     =
         \overline{\mu}_k \nabla \cdot ( \overline{\sf{D}}_k \cdot\nabla \overline{\mu}_k) 
    +          \overline{\mu}_k \nabla \cdot \Big( \overline{\sf{D}}_k \cdot\nabla \big<\mu_k^\prime \big > \Big) \nonumber \\ 
    +  \overline{\mu}_k \Big<\nabla \cdot ( \sf{D}_k^\prime \cdot\nabla \mu_k) \Big >     
    + \Big< \mu_k^\prime  \nabla \cdot ( \sf{D}_k \cdot\nabla \mu_k) \Big > \;.
\nonumber
\end{eqnarray}
Scale invariance for Fick's law will therefore be obtained if the following conditions are satisfied
\begin{eqnarray}
     \big<\nabla \mu_k^\prime \big >   &=& {0} \;,   \label{eq:fick-condition-a} \\
    \Big<\nabla \cdot ( \sf{D}_k^\prime \cdot\nabla \mu_k) \Big >      &=& 0 \;,   \label{eq:fick-condition-b} \\
    \Big< \mu_k^\prime  \nabla \cdot ( \sf{D}_k \cdot\nabla \mu_k) \Big >  &=& 0\;,    \label{eq:fick-condition-c}
\end{eqnarray}
where the deviation for the diffusion tensor is $\sf{D}_k^\prime=\sf{D}_k -\overline{\sf{D}}_k$.
In ideal systems the chemical potential is directly related to the density,
\begin{equation}
    \mu = \mu_0 + RT \ln (\rho / \rho_0)\;,
    \label{eq:ideal-gas-potential}
\end{equation} with $\mu_0$ and $\rho_0$ being constant
reference values. Ideal conditions imply that Eq. \ref{eq:fick-condition-a} must hold due to the definition of $\overline{\rho}_k$. Eq. \ref{eq:fick-condition-b} will hold if $\sf{D}_k$ is independent of space
and time, since this would mean that $\sf{D}_k^\prime = 0$.  
Finally,  Eq. \ref{eq:fick-condition-c} suggests that the existence of gradients in $\sf{D}_k$ or $\mu_k$
that have a length scale smaller than $\Omega$ will cause Fick's law to fail at the averaged scale, 
as will transient changes to these gradients that occur faster than the averaging timescale $\Lambda$. At the intuitive level, these criteria require that the diffusion coefficient, composition and chemical potential should vary only linearly over $\Lambda$ and $\Omega$. For such cases the fluctuation terms disappear and Fick's law will hold in the larger-scale system. However, in real systems non-Fickian behaviour is quite common \cite{Battiato_etal_2009,Battiato_Tartakovsky_2011,Cushman_OMalley_2015}.
Diffusive and dispersive processes within complex micro-structure give rise to the development of heterogeneous structures, including fractals, based on local instabilities and material heterogeneity \cite{Cushman_Ginn_1993}. The implication is that Eqs. \ref{eq:fick-condition-a} and \ref{eq:fick-condition-b} do not hold under typical conditions, since mass transport routinely leads to transient spatial heterogeneity for the composition, chemical potential, and diffusion coefficient.

\subsection*{Capillary fluctuations}

In heterogeneous systems it is common for chemical potential gradients to exist at multiple length scales. 
Multiphase systems arise due to the dependence of $\mu$ on composition, e.g. in the Cahn-Hilliard equations,
\begin{equation}
    \mu = \phi^3 - \phi - \gamma \nabla^2 \phi \;.
    \label{eq:Cahn-Hilliard}
\end{equation}
The order parameter $\phi = \rho-(\rho_l^*+\rho_g^*)/2$, 
where the equilibrium densities for the liquid and gas are
$\rho_l^*$ and $\rho_g^*$. The parameter $\gamma$ determines the width 
of the interface separating the two pure phase regions. It is entirely possible to describe 
multiphase systems using Eq. \ref{eq:Cahn-Hilliard} in combination with Eq. \ref{eq:entropy-inequality-mass}.
As a matter of convenience, we can also describe the thermodynamics of the system
by considering Eq. \ref{eq:euler-homogeneous} in the form
\begin{equation}
    U = U(S,V_w,V_n,A_{wn})\;,
    \label{eq:capillary-energy}
\end{equation}
where $V_w$ and $V_n$ are the volume of the droplet and the surrounding fluid and
$A_{wn}$ is the interfacial area between fluids. The associated intensive variables
are the pressures $p_w$ and $p_n$ and the interfacial tension $\gamma_{wn}$. 
Averaging in time and space we assume: (1) isothermal conditions; 
(2) constant interfacial tension;
(3) the volume of each fluid is constant; and
(4) compositional effects are negligible.
Subject to these restrictions Eq. \ref{eq:entropy-inequality} simplifies to
\begin{equation}
   \frac{\partial\overline{S}}{\partial t}  = -\frac {1}{\overline{T}} \Big[
    \Big< V_w \frac{\partial{p}_w^\prime}{\partial t}\Big> 
        + \Big< V_n \frac{\partial{p}_n^\prime}{\partial t}\Big> 
    + \overline{\gamma}_{wn} \frac{\partial\overline{A}_{wn}}{\partial t} 
    \Big] \ge 0 \;.
    \label{eq:entropy-inequality-droplet}
\end{equation}
Dissipative effects are understood by considering the fluctuation of the fluid pressures that are induced by the spontaneous change in surface energy. 

\begin{figure}[ht]
\centering
\includegraphics[width=1.0\linewidth]{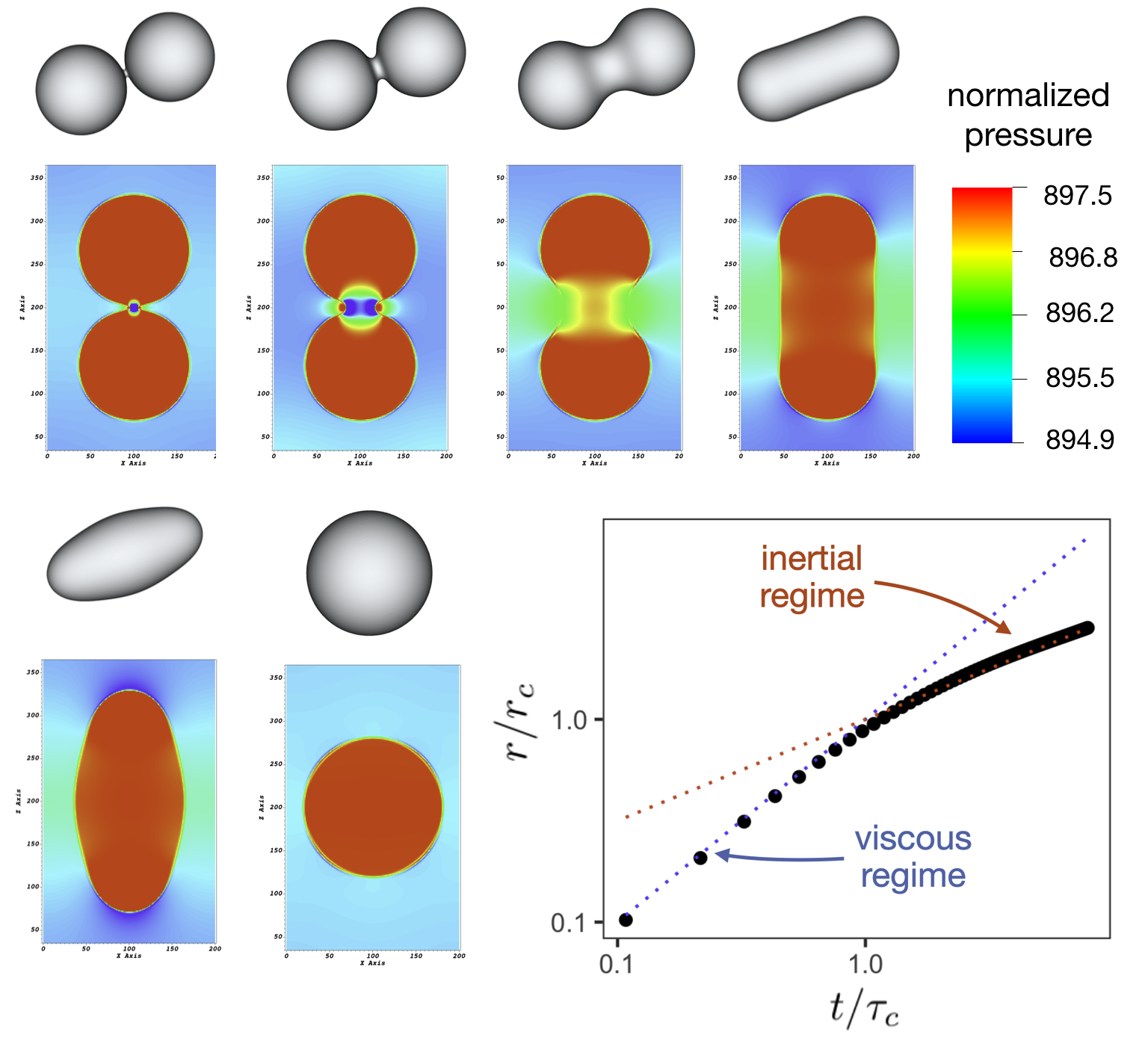}
\caption{Droplet coalescence leads to fluctuations in the
pressure field (color) due to the rapid change in capillary forces. 
Scaling for the coalescence event is limited by viscous forces during the 
initial bridge formation, and is dominated by inertial forces at late times.}
\label{fig:droplet-coalescence}
\end{figure}

The coalescence of two fluid droplets represents one the simplest examples of topological change in fluid mechanics. The topological event induces an apparent singularity followed by a cascade of energy dissipation governed by distinct scaling regimes as the system establishes a new equilibrium. 
At equilibrium, the fluid pressures and interface curvature are related based on the Young-Laplace equation 
\begin{equation}
    p_n - p_w = \gamma_{wn} \Bigg( \frac 1{R_1} + \frac 1{R_2}\Bigg)\;,
    \label{eq:young-laplace}
\end{equation}
where $R_1$ and $R_2$ are the principal curvatures along the interface between fluids. 
Detailed studies of the droplet coalescence mechanism show that the flow behavior 
and associated geometric evolution are coupled on a very fast timescale \cite{Pauslon_NatureComm_2012,Dirk_2005,Ristenpart_2006,Wu_2004,Orme_1997,Pauslon_PRL_2008,Pauslon_PRL_2011,Pauslon_PNAS_2012}. At the molecular level, coalescence is initiated based on thermal 
effects \cite{Perumananth_etal_PRL_2019}. Hydrodynamic effects become dominant after approximately 30 picoseconds, 
based on the formation of a bridge that joins the droplets. The ensuing dynamics can be separated into two distinct regimes
\cite{Pauslon_PRL_2011}. 
At early times, viscous effects dominate and the growth of the bridge radius scales as
\begin{equation}
     \frac{r}{r_c} \sim \frac{t}{\tau_c}\;,
          \label{eq:coalescence-scaling-i}
 \end{equation}
where $\tau_c$ is the crossover time and $r_c$ is the associated length scale.
At late times, inertial effects dominate and the growth of the bridge radius scales as
 \begin{equation}
     \frac{r}{r_c} \sim \sqrt{\frac{t}{\tau_c}}\;.
     \label{eq:coalescence-scaling-ii}
 \end{equation}
Analogous results have been obtained for droplet snap-off 
\cite{Pahlavan13780}. The rate of entropy production will be a non-linear function
of time based on the crossover between the two regimes. The non-equilibrium
response is therefore non-linear. Averaging in time and space can be applied to 
homogenize these non-linear dynamics so that they can be treated explicitly
as energy fluctuations. 

The sequence depicted in Fig. \ref{fig:droplet-coalescence}
shows the effect of the coalescence event on the fluid pressure field as simulated
by a lattice Boltzmann model \cite{mcclure2020lbpm}. Results demonstrate that simulation accurately recovers the predicted scaling
behavior. In Fig. \ref{fig:droplet-coalescence} the pressure field $p_i^*=p_i R_{f} /\gamma_{wn}$,
is normalized relative to the final droplet radius, $R_f=80$ voxels.
Non-equilibrium effects develop in response to the near instantaneous curvature disruption
along the interface at the point where the droplets first touch. The associated pressure shock drives 
the ensuing dynamics. The behavior is non-ergodic because capillary energy barriers inhibit the thermal mixing between the droplets prior to the event. Rapid mixing occurs after coalescence, once the energy barrier separating the two droplets has been destroyed. 

Thermal fluctuations due to molecular effects are usually assumed to be stationary with respect to 
time as a consequence of the time reversibility of the Hamiltonian at the molecular level 
\cite{Kubo_1957}. Capillary fluctuations are distinct from thermal fluctuations due to the fact that they are inherently cooperative in nature, are linked to both reversible and irreversible energy transfer.
Symmetry-breaking is a consequence of the structural rearrangement associated with the transition to a new local minimum energy configuration. For a closed system the dissipated energy is easily calculated from the initial and final configurations, since at equilibrium the droplets are spherical,
\begin{eqnarray}
 \Delta S = \frac 1T \big( V_w \Delta p_w + V_n \Delta p_n - \gamma_{wn} \Delta A_{wn} \big)\;.
 \label{eq:droplet-thermodynamics-i}
\end{eqnarray}
The radius for the initial and final droplets can be computed analytically based on the volume, making use of the 
Young-Laplace equation to determine the equilibrium fluid pressures. 

\begin{figure*}[ht]
\centering
\includegraphics[width=0.9\linewidth]{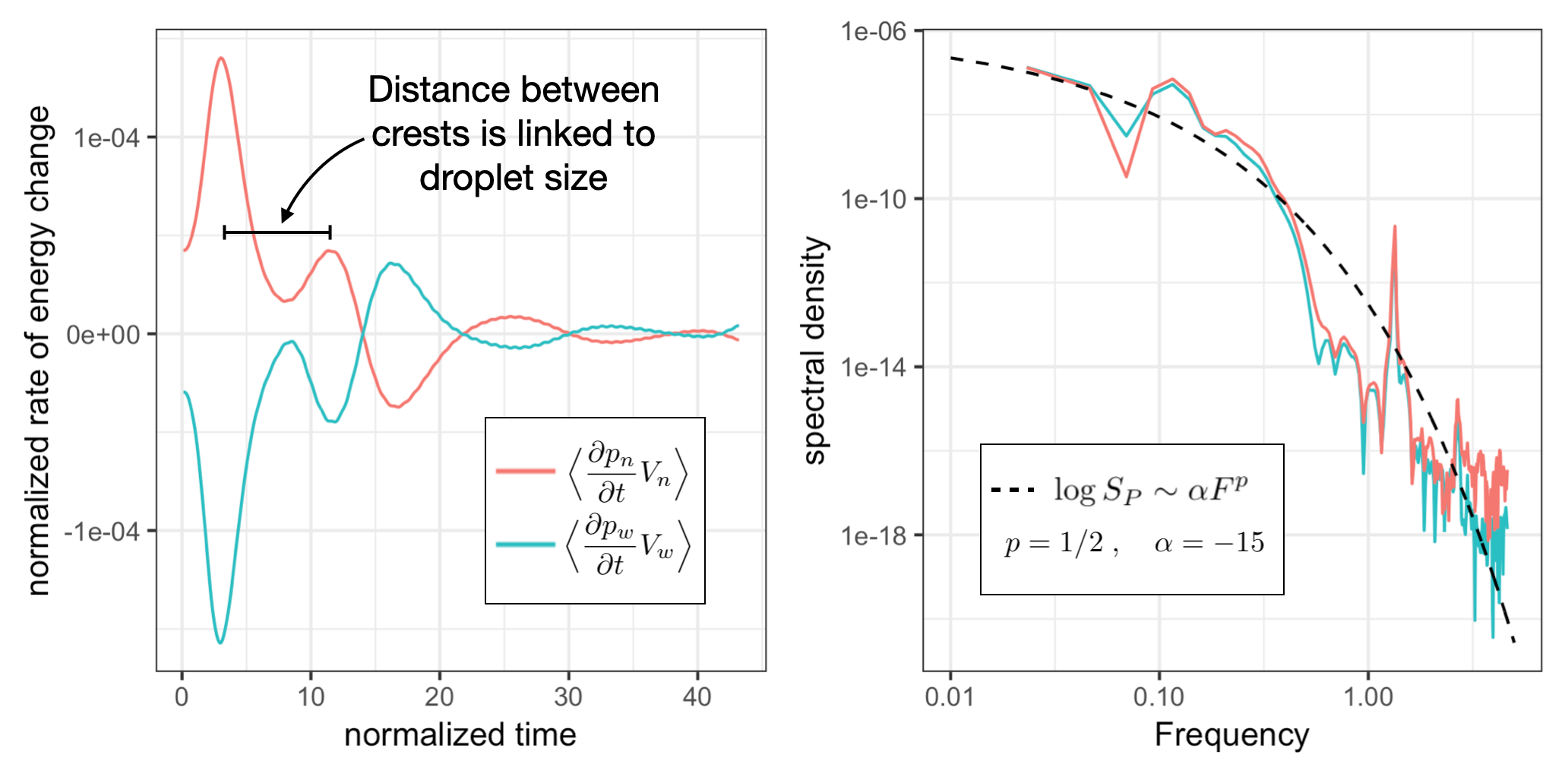}
\caption{Pressure fluctuation due to the coalescence of two fluid droplets.
(A) rate of energy change associated with pressure fluctuation relative to the final surface energy during droplet coalescence event; and
(B) noise spectrum associated with capillary fluctuations can be predicted by $\log S_P \sim \alpha F^p$.
}
\label{fig:coalescence-energy}
\end{figure*}

The rate of energy change associated with capillary fluctuations is shown in Figure \ref{fig:coalescence-energy}. The timescale is normalized by the crossover time, $\tau_c$, which can be considered as defining the intrinsic timescale for coalescence.
In our analysis, the full system is used as the domain for spatial averaging and the time interval is $\Lambda=0.108\tau_c$. Since this timescale is faster than the non-linear dynamics, the frequency for the fluctuations is directly visible in Figure
\ref{fig:coalescence-energy}a. The energy scale is normalized based on the interfacial tension and the capillary length scale, which is accomplished by
dividing the energy associated with capillary fluctuations final surface energy $4 \pi \gamma_{wn} R_f^2$. The
scales in Fig. \ref{fig:coalescence-energy} are therefore non-dimensional.
From  Eq. \ref{eq:entropy-inequality-droplet} it is clear that the changes in surface energy are driving the fluctuations. 
Immediately after coalescence, interfaces perform local work against the fluid
pressure due to unbalanced capillary forces. The pressure fluctuation terms are understood as resulting from local gradients in the pressure field that are generated due to the curvature discontinuity. 
These gradients are clearly visible in Fig. \ref{fig:droplet-coalescence}.
Since the initial and final droplet states have different capillary pressure, the pressure fluctuation is not stationary over the event. Nevertheless, a degree of symmetry is clear based on the mirroring effect between the pressure fluctuation in one fluid and the other. This suggests that the choice to represent the system in a discrete way introduces asymmetry into the system description. The Gibb's dividing surface sub-divides the system into distinct sub-regions for each fluid, each with its own pressure. While topological changes such as droplet coalescence are fundamentally discrete events when considered from the perspective of set theory, set construction is a choice imposed on the system, as opposed to an underlying property of the physical system itself. Set operations are the basis for separately defining $p_w$ and $p_n$. 
Detailed balance is not evident when the fluids are considered separately,
since irreversible energy exchanges occur between the fluids due to cooperative rearrangement of the interface. 
For this reason symmetry is a property of the global system and not a property of the sub-regions. 

Pressure fluctuations are multiscale rate effects that arise due to
spatial heterogeneity. The length scale associated with the heterogeneity
is the driving factor that determines the spectral properties. Due to length scale effects 
the noise signature associated with capillary fluctuations is distinct from pink noise, 
where the relationship between the spectral density $S_P$ and frequency $F$ is
$S_P \sim 1/F$ \cite{Kendal_2015,Bak_1987,Weissman_1988}. For the data shown in
Fig. \ref{fig:coalescence-energy}, the time and energy scales are non-dimensionalized
as described previously. A simple power law is insufficient to describe noise due to capillary fluctuations.
Instead, we find that the scaling relationship is a stretched exponential,
\begin{equation}
    \log S_P \sim \alpha F^p  \;.
    \label{eq:spectrum}
\end{equation}
The coefficients $\alpha$ and $p$ are associated with the rate of decay in the spectral density as the frequency increases.  $S_P$ decays rapidly with $F$ for frequencies that are faster than a typical event.
This can be considered as a transition between two distinct scaling regimes. At the length and time scale
for the coalescence event the behavior is super-diffusive based on the fact that cooperative capillary forces 
move mass faster than the local diffusion rate. Ultimately the capillary forces that drive these
events originate due to gradients in composition at smaller scales. At smaller length scales sub-diffusive behavior is 
obtained based on the strong anti-diffusion associated with the interface region. 
The crossover between these distinct scaling regimes leads to a corresponding transition in the fluctuation spectrum.
For droplet coalescence $\alpha=-15$ and $p=1/2$ match well with the simulated fluctuation spectrum.

\begin{figure*}[ht]
\centering
\includegraphics[width=0.9\linewidth]{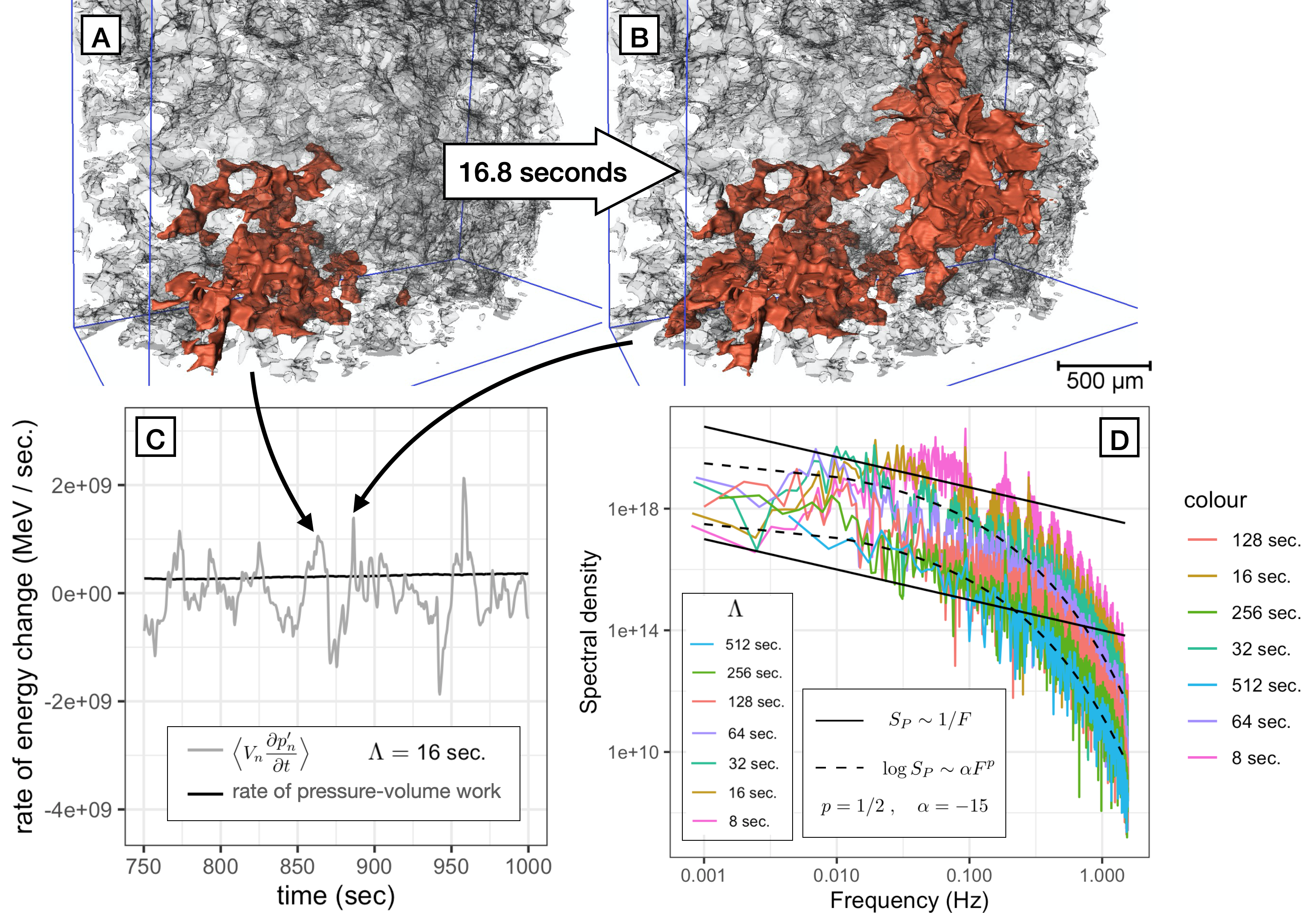}
\caption{Capillary fluctuations during displacement in porous media
are linked with spontaneous pore-scale events that occur due to capillary
forces within the solid micro-structure. Haines jumps occur spontaneously
during immiscible displacement as the system ``jumps" from (A)  one energy minimum 
to (B) the next; (C) fluid pressure fluctuations arise due to these dynamics; and
(D) scaling behavior for the power spectrum is independent of the time averaging interval.
}
\label{fig:Berea-fluctuation}
\end{figure*}

The scaling relationship given by Eq. \ref{eq:spectrum} also holds for immiscible displacement in porous media. 
Cooperative events occur routinely as fluids migrate through complex micro-structure under the influence of capillary forces \cite{Morrow_1970}. Experimental data demonstrates that the timescale for these events is directly linked to the frequency for fluctuations in the pressure signal \cite{Alpak_2019}. Results shown in Fig. \ref{fig:Berea-fluctuation} were obtained using synchrotron micro-tomography imaging. For complete experimental details the reader is referred to Berg et al. \cite{Berg_Ott_etal_13}. The system was initially saturated with brine, and oil was injected into the sample at a rate of $0.35$ $\mu$L/minute. Pressure transducer measurements were collected at an interval $\Delta t = 0.32$ sec. 
The absolute permeability for the Berea sandstone was $\kappa = 0.7$ $\mu\mbox{m}^2$. 
Initial and final states for a pore-scale event known as a Haines jump are shown in Fig. \ref{fig:Berea-fluctuation}A and \ref{fig:Berea-fluctuation}B. Haines jumps are spontaneous events that occur when the fluid
meniscus passes through narrow pore throats within the solid micro-structure, where the capillary pressure is high. 
As fluid spontaneously flows into the adjacent pore body, the capillary pressure drops rapidly causing a fluctuation 
in the signal. The timescale for pore-scale events is controlled by the solid micro-structure, which means that the statistics for the associated noise signal are linked to length scale heterogeneity within the system. 
In spite of this considerable complexity, the scaling behavior is predicted from Eq. \ref{eq:spectrum} with identical coefficients to the droplet coalescence. In the spectral analysis, a non-dimensional
timescale for the experiments can be defined as
\begin{equation}
    t^* = \frac{\gamma_{wn} t}{\mu \sqrt{\kappa}} \;,
\end{equation}
where $\mu=0.89$ ${\mbox{mN} \cdot \mbox{s}} / \mbox{m}^2 $ and $\gamma_{wn} = 25$ $\mbox{mN} / \mbox{m}$. 
Since the permeability is related to the microscopic length scale, a non-dimensional energy scale can be defined
by normalizing based on $\gamma_{wn} \kappa$. 
Averaging over a longer time interval reduces the amplitude of the fluctuations, 
but does not change $\alpha$ and $p$.
Both droplet coalescence and Haines jump events are examples of critical phenomena. The length scale
for an individual Haines jump is controlled by the solid microstructure, and a single event may cause fluids
to invade multiple pores very rapidly, as shown in Fig. \ref{fig:Berea-fluctuation}A--B. The timescale
for pressure fluctuations is therefore determined based on the distribution of length scales for the solid
microstructure, which is typically very heterogeneous. It should be expected that the distribution of fluids
will exert strong influence on the fluctuation spectrum, as the probability for fluid coalescence events 
depends on how mass is distributed within the solid micro-structure. Further study is needed to understand
if universal coefficients can be established in heterogeneous systems. 
 As in droplet coalescence, the pressure fluctuations are non-stationary. 
For the data shown in Fig. \ref{fig:Berea-fluctuation} the net contribution from fluctuations is $7.84$\% of the total pressure-volume work when considered over $\Lambda=512$ seconds. In this situation
fluctuation terms must be included explicitly in the non-equilibrium thermodynamics, since these
terms are needed to state conservation of energy for the system. 

\subsection*{Summary and Conclusions}

We derive non-equilibrium thermodynamic expressions using time-and-space averaging, 
showing that fluctuations occur due to non-linear dynamics associated with the intensive variables. Considering a particular timescale of interest, the approach is constructed to treat systems that are ergodic at small length scales but non-ergodic at larger scales. Time and space averages are formulated by directly integrating
the energy dynamics, meaning that the actual system evolution is  
captured within the averaged representation. The approach is scale-consistent based on the fact that
thermodynamic quantities retain their physical meaning, and the form of the Euler equation will independent
of the length and time scales. Fluctuations are linked with sub-scale gradients in systems that exhibit multiscale heterogeneity. When stimulated, these systems have a tendency to relax toward equilibrium at  
macroscopically slow timescales determined based on the length scale associated with the gradients. 
We illustrate how anomalous diffusion can arise in a system where Fick's law applies
at some small length scale, but fails at larger scales due spatial heterogeneity. 
Explicit conditions for scale invariance are obtained. 

Averages in time demonstrate that the rate of entropy production is scale dependent, 
e.g. due to crossover times that are frequently encountered when considering the dynamic response
of heterogeneous systems. For any stationary process, the time average of the rate of entropy 
production is zero, since net changes to any thermodynamic state function imply that a system is 
non-stationary. Averaging does not alter the interpretation for the Clausius theorem, since 
averages are constructed based on conservation of energy. Treatment of stationary processes is 
thereby simplified, since dissipation is expressed in terms of the rate of work and heat
exchange, which are linked to fluctuations based on the internal energy dynamics of the system. 
We consider basic applications to the Carnot cycle and to Maxwell's demon to illustrate how thermodynamic 
cycles can can be understood in the context of fluctuations.

More generally, fluctuations describe the internal energy dynamics
of thermodynamic systems away from equilibrium. Since fluctuations must conserve energy, symmetry laws 
can be derived relating fluctuations within heterogeneous systems. These statements govern the possible 
energy transfers that can occur within non-equilibrium systems. In contrast with thermal fluctuations, multiscale fluctuations are linked with cooperative events and therefore may not obey 
detailed balance. Non-stationary fluctuations occur when there are net energy transfers
within a system. We consider droplet coalescence as a clear example of a non-equilibrium
system where multiple crossover times are present. In contrast with a thermodynamic cycle,
droplet coalescence is non-stationary based on the fact that entropy production occurs. 
Considering coalescence along
with immiscible displacement in porous media, we show that capillary fluctuations can be predicted by a simple scaling law, $\log S_P \sim \alpha F^p$, with $\alpha=-15$ and $p=1/2$ matching the fluctuation spectrum in both cases.
Multiscale fluctuations are extensible to other systems where gradients in composition, chemical potential, and other intensive variables are operative at a scale smaller than the scale of interest. Classes of non-ergodic behavior that are defined by macroscopically slow physics are particularly relevant. The formulation is based on classical thermodynamic theory, and defines fluctuations in terms of standard quantities that are straightforward to measure in a practical setting.

\subsection*{Acknowledgements}
\begin{acknowledgments}
`An award of computer time was provided by the Department of Energy Summit Early Science program. This research also used resources of the Oak Ridge Leadership Computing Facility, which is a DOE Office of Science User Facility supported under Contract DE-AC05-00OR22725. $\mu$CT was performed on the TOMCAT beamline at the Swiss Light Source, Paul Scherrer Institut, Villigen, Switzerland. We are grateful to G. Mikuljan at Swiss Light Source, whose outstanding efforts have made these experiments possible'
\end{acknowledgments}


\begin{thebibliography}{10}

\bibitem{parry2004topics}
W.~Parry,
\newblock {\em Topics in Ergodic Theory}.
\newblock Cambridge Tracts in Mathematics. Cambridge University Press, 2004.

\bibitem{Neumann70}
J.~v. Neumann,
\newblock {Proc. Natl. Acad. Sci. U.S.A.} {\bf18} 70 (1932).

\bibitem{Birkhoff656}
G.~D. Birkhoff,
\newblock {Proc. Natl. Acad. Sci. U.S.A.} {\bf 17}, 656 (1931).

\bibitem{Palmer_1982}
R.G. Palmer,
\newblock {Adv. Phys.} {\bf 31}, 669 (1982).

\bibitem{Weigel6438}
A~V. Weigel, B. Simon, M.~M. Tamkun and D. Krapf,
\newblock {Proc. Natl. Acad. Sci. U.S.A.} {\bf 108}, 6438, 2011.

\bibitem{Schwarzl_2017}
M. Schwarzl, A. Godec and R. Metzler,
\newblock { Sci. Reports}, {\bf 7}, 3878 (2017).

\bibitem{Gnesotto_2018}
F.~S. Gnesotto, F.~Mura, J.~Gladrow and C.~P. Broedersz,
\newblock { Reports Prog. Phys.}, {\bf 81}, 066601 (2018).

\bibitem{Debenedetti_2001}
P.~G. Debenedetti and F.~H. Stillinger,
\newblock { Nature}, {\bf 410}, 159 (2001).

\bibitem{Crisanti_2003}
A.~Crisanti and F.~Ritort,
\newblock { J. Phys. A: Math. Gen.} {\bf 36}, R181 (2003).

\bibitem{Tarjus_2005}
G.~Tarjus, S.~A. Kivelson, Z.~Nussinov and P.~Viot,
\newblock {J. Phys. Cond. Matter}, {\bf 17}, R1143 (2005).

\bibitem{PhysRevLett.111.064501}
S.S. Datta, H.~Chiang, T.~S. Ramakrishnan and D.~A. Weitz,
\newblock {Phys. Rev. Lett.}, {\bf 111}, 064501, (2013).

\bibitem{PhysRevE.88.052116}
V.~D. Borman, A.~A. Belogorlov, V.~A. Byrkin and V.~N. Tronin.
\newblock { Phys. Rev. E}, {\bf88}, 052116, (2013).

\bibitem{CRANDALL2009574}
D. Crandall, G. Ahmadi, M. Ferer and D.~H. Smith,
\newblock { Physica A: Stat. Mech. App.}, {\bf 388}, 574 (2009).

\bibitem{PhysRevLett.87.055701}
E.~Kierlik, P.~A. Monson, M.~L. Rosinberg, L.~Sarkisov and G.~Tarjus,
\newblock { Phys. Rev. Lett.}, {\bf 87} 055701 (2001).

\bibitem{Winkler_etal_2019}
M. Winkler, M. Gjennestad, D. Bedeaux, S. Kjelstrup, R. Cabriolu and A. Hansen,
\newblock{Frontiers Phys.}, {\bf 8},  {60} {(2020)}.      

\bibitem{Iwamatsu_nucleation_2011}
M. Iwamatsu.
\newblock {J. Chem. Phys.} {\bf 134} 164508 (2011).

\bibitem{RevModPhys.90.015006}
A. Baule, F. Morone, H.~J. Herrmann and H.~A. Makse,
\newblock {Rev. Mod. Phys.}, {\bf 90}, 015006 (2018).

\bibitem{Herrmann1998}
H.J. Herrmann and S.~Luding,
\newblock {Continuum Mech.  Thermodynamics}, {\bf 10}, 189 (1998).

\bibitem{Mehta8244}
A.~Mehta, G.~C. Barker and J.~M. Luck,
\newblock {Proc. Natl. Acad. Sci. U.S.A.},
 {\bf 105}, 8244 (2008).

\bibitem{Atia_etal_2018}
L. Atia, D. Bi, Y. Sharma, J.~A. Mitchel, B. Gweon, S.
  A.~Koehler, S.~J. DeCamp, B.~Lan, J.~H. Kim, R. Hirsch, A.~F.
  Pegoraro, K.~H. Lee, J.~R. Starr, D.~A. Weitz, A.~C. Martin,
  J.-A. Park, J.~P. Butler and J.~J. Fredberg,
\newblock { Nature Phys.} {\bf14}, 613 (2018).

\bibitem{Callen_1951}
H.~B. Callen and T.~A. Welton.
\newblock { Phys. Rev.}, {\bf 83}, 34 (1951).

\bibitem{Kubo_1957}
R. Kubo,
\newblock {J. Phys. Soc. Japan}, {\bf 12}, 570 (1957).

\bibitem{PhysRevLett.103.090601}
J.~Prost, J.-F. Joanny, and J.~M.~R. Parrondo,
\newblock {Phys. Rev. Lett.}, {\bf 103}, 090601, (2009).

\bibitem{Maes_2020}
C. Maes,
\newblock {Phys. Rev. Lett.}, {\bf 125}, 208001 (2020).

\bibitem{Wolpert_2020}
D.~H. Wolpert,
\newblock {Phys. Rev. Lett.}, {\bf 125}, 200602, (2020).

\bibitem{Morrow_1970}
N.~R. Morrow,
\newblock {Indust. Eng. Chem.}, {\bf62}, 32 (1970).

\bibitem{Armstrong_Berg_2013}
 {R. T. Armstrong and S. Berg},
 \newblock{Phys. Rev. E}, \textbf{88}, {043010}, (2013).

\bibitem{Berg_Ott_etal_13}
S. Berg, H. Ott, S.A. Klapp, A. Schwing, R. Neiteler, N.
  Brussee, A. Makurat, L. Leu, F. Enzmann, J. Schwarz,
  M. Kersten, S. Irvine,  and M. Stampanoni,
\newblock {Proc. Natl. Acad. Sci. U.S.A.}, {\bf110}, 3755, (2013).

\bibitem{Armstrong_Ott_etal_2014}
R.T. Armstrong RT, H. Ott, A. Georgiadis, M. R\"{u}cker, A. Schwing and S. Berg, 
\newblock{Water Resour. Res.} {\bf 50} 9162 (2014).


\bibitem{Cueto-Felgueroso_Juanes_2015}
L. Cueto-Felgueroso and R. Juanes,
\newblock{Geophys. Res. Lett.} {\bf 43}, 1615 (2015)

\bibitem{Morrow_1970}
 N.R. Morrow,
\newblock{Ind. Eng. Chem} {\bf 63}, 32 (1970).

\bibitem{Hassanizadeh_Gray_93}
S.M. Hassanizadeh and W.G. Gray,
\newblock { Adv. Water Res.}, {\bf16}, 53 (1993).

\bibitem{Bear_Nitao_1995}
 J. Bear and J.J. Nitao,  
\newblock{Transp. Porous Med.} { \bf18,} 151 (1995).

\bibitem{marle1981multiphase}
  {C. Marle},
  \newblock{``Multiphase Flow in Porous Media"},
{Editions Technip} {(1981)}.

\bibitem{Aryana_2013}
S.A. Aryana and A.R. Kovscek, 
\newblock{Transp. Porous Med.} {\bf 97}, 373 (2013). 


\bibitem{Primkulov_2019}
B. K. Primkulov, A.A. Pahlavan, X. Fu, B. Zhao, C.W. MacMinn and R. Juanes,  
\newblock{J. Fluid Mech.} {\bf 875}, R4 (2019).

\bibitem{Pahlavan13780}
A.~A. Pahlavan, H.~A. Stone, G.~H. McKinley and R. Juanes.
\newblock {Proc. Natl. Acad. Sci. U.S.A.} {\bf116}, 13780 (2019).

\bibitem{callen1960thermodynamics}
H.B. Callen.
\newblock {\em Thermodynamics}, (1960).

\bibitem{Grmela_1997a}
M. Grmela and H.C. \"{O}ttinger,
\newblock {Phys. Rev. E.} {\bf56}, 6620 (1997). 

\bibitem{Grmela_1997b}
M. Grmela and H.C. \"{O}ttinger,
\newblock {Phys. Rev. E.} {\bf56}, 6633 (1997). 

\bibitem{Gray_Miller_14}
W.G. Gray and C.T. Miller,
\newblock {{I}ntroduction to the {T}hermodynamically {C}onstrained {A}veraging {T}heory for {P}orous {M}edium {S}ystems.}
Springer, Z\"{u}rich, (2014).

\bibitem{Froemberg_2013}
    {D. Froemberg and E. Barkai},
     \newblock{Phys Rev E}, {\bf 88}, {2}    (2013).

\bibitem{degroot_Mazur_2013}
S.R. De~Groot and P.~Mazur,
\newblock {\em Non-Equilibrium Thermodynamics}, Dover (2013).

\bibitem{maxwell_1871}
James~Clerk Maxwell.
\newblock {\em Theory of Heat}.
\newblock Cambridge Library Collection - Physical Sciences. Cambridge
  University Press, 2011.
  
\bibitem{maxwell_1879}
\newblock{Nature} 20, 126 (1879). 

\bibitem{Earman_Norton_1999}
J. Earman, J. D. Norton,
\newblock{Studies in History and Philosophy of Science Part B: Studies in History and Philosophy of Modern Physics},
Volume 30, 1 (1999).

\bibitem{Ben-Naim_2008}
A. Ben-Naim,
\newblock{A Farewell to Entropy},
{(World Scientific)},({2008}).

\bibitem{Sackur_1911}
O. Sackur
\newblock{Annalen der Physik}, {\bf 36}, 958 (1911).

\bibitem{Tetrode_1912}
H. Tetrode 
\newblock{Annalen der Physik} {\bf 38}, 434 (1912).

\bibitem{Shiraishi_2015}
{N. Shiraishi, Naoto and T. Sagawa},
\newblock {Phys. Rev. E}, {\bf 91}, {012130} (2015).

\bibitem{Rupprecht_2019}
N. Rupprecht and D. Can Vural,
\newblock {Phys. Rev. Lett.} 123, 080603 (2019).

\bibitem{Rupprecht_2020}
N. Rupprecht and D. Can Vural,
\newblock{Phys. Rev. E}, {\bf 102}, 062145 (2020).

\bibitem{Koski_2014}
J.V. Koski, V.F. Maisi, T. Sagava and J.P. Pekola, 
\newblock{Phys. Rev. Lett.} {\bf113} 030601 (2014).

\bibitem{Touchette_2000}
 Hugo Touchette and S. Lloyd 
\newblock{Phys. Rev. Lett.} {\bf 84} 1156 (2000). 

\bibitem{Sagawa_2010}
 T. Sagawa and M. Ueda,
 \newblock{Phys. Rev. Lett.} {\bf 104} 090602.  (2010).
 
\bibitem{Onsager_1931a}
L. Onsager, 
\newblock{Phys. Rev.} {\bf 37}, 405 (1931).

\bibitem{XU2006452}
J. Xu, S. Kjelstrup, D. Bedeaux, A. Røsjorde, L. Rekvig
\newblock{Journal of Colloid and Interface Science} {\bf 299}, 452 (2006).

\bibitem{Battiato_Tartakovsky_2011}
I. Battiato, D.M. Tartakovsky,
\newblock{Journal of Contaminant Hydrology},
{\bf120–121}, 18  (2011).

\bibitem{Battiato_etal_2009}
I. Battiato, D.M. Tartakovsky, A.M. Tartakovsky, T. Scheibe,
\newblock{Advances in Water Resources}, {\bf32}, 1664 (2009).

\bibitem{Cushman_OMalley_2015}
J. H. Cushman and D. O’Malley,
\newblock{Journal of Hydrology},
{\bf 531}, 161 (2015).

\bibitem{Cushman_Ginn_1993}
J.H. Cushman and T.R.  Ginn, 
\newblock{Transp Porous Med} {\bf 13}, 123 (1993).

\bibitem{Noether_1918}
E.~Noether,
\newblock { Gott. Nachr.}, {\bf 235}, 405 (1918).

\bibitem{Noether_1971}
E.~Noether,
\newblock {Transport Theory Stat. Phys.} {\bf1}, 186 (1971).

\bibitem{Pauslon_NatureComm_2012}
J.D. Paulsen, R.~Carmigniani, A. Kannan, J.~C. Burton and S.~R. Nagel,
\newblock { Nat. Commun.}, {\bf 5}, 3182 (2014).

\bibitem{Dirk_2005}
G.~A.~L. Dirk, H.N. Aarts, H.N.~W. Lekkerkerker, H. Guo, G.H. Wegdam and D.~Bonn,
\newblock { Phys. Rev. Lett.}, {\bf 95}, 164503 (2005).

\bibitem{Ristenpart_2006}
W.~D. Ristenpart, P.~M. McCalla, R.~V. Roy and H.~A. Stone,
\newblock {Phys. Rev. Lett.}, {\bf 97}, 064501 (2006).

\bibitem{Wu_2004}
M.~Wu,
\newblock {Phys. Fluids.}, {\bf16}, L51 (2004).

\bibitem{Orme_1997}
M.~Orme,
\newblock {Progress in Energy and Combustion Sci.}, {\bf 23}  65 (1997).

\bibitem{Pauslon_PRL_2008}
S.~C. Case and S.~R. Nagel,
\newblock { Phys. Rev. Lett.}, {\bf 100}, 084503 (2008).

\bibitem{Pauslon_PRL_2011}
J.D. Paulsen, J.C. Burton and S.R. Nagel,
\newblock { Phys. Rev. Lett.}, {\bf 106}, 114501 (2011).

\bibitem{Pauslon_PNAS_2012}
J.D. Paulsen, J.C. Burton, S.R. Nagel, S.~Appathurai, M.T. Harris and O.A. Basaran,
\newblock {Proc. Natl. Acad. Sci. U.S.A.}, {\bf 109}, 6859 (2012).

\bibitem{Perumananth_etal_PRL_2019}
S. Perumanath, M.~K. Borg, M.~V. Chubynsky, J.~E. Sprittles and J.~M. Reese,
\newblock { Phys. Rev. Lett.}, {\bf 122}, 104501 (2019).

\bibitem{mcclure2020lbpm}
J.~E. McClure, Z. Li, M. Berrill and T. Ramstad,
arXiv:2007.12266, (2020).

\bibitem{Weissman_1988}
 M. B.  Weissman,
\newblock{Rev. Mod. Phys.} {\bf 60}, 537 (1988). 

\bibitem{Kendal_2015}
W.S. Kendal, 
\newblock{Physica A} {\bf 421}, 141 (2015).

\bibitem{Bak_1987}
P. Bak, C. Tang, and K. Wiesenfeld,
 \newblock{Phys. Rev. Let.} {\bf 59}, 381 (1987). 
 
 \bibitem{Alpak_2019}
F. Alpak, I Zacharoudiou, S Berg, J Dietderich, N Saxena,
\newblock{Computational Geosciences}, {\bf 23}, 849--880 (2019)

\end{thebibliography}

\end{document}